\documentclass[conference]{IEEEtran}
\IEEEoverridecommandlockouts
\usepackage{cite}
\usepackage{amsmath,amssymb,amsfonts}
\usepackage{algorithmic}
\usepackage{graphicx}
\usepackage{textcomp}
\usepackage{xcolor}
\usepackage{booktabs}
\usepackage{url}
\usepackage[hidelinks]{hyperref}
\def\BibTeX{{\rm B\kern-.05em{\sc i\kern-.025em b}\kern-.08em
    T\kern-.1667em\lower.7ex\hbox{E}\kern-.125emX}}
\begin{document}

\title{Hybrid Neural Retrieval with Generative Query Refinement for Quranic Passage Retrieval \\
}
\author{
\IEEEauthorblockN{Mohamed G. Salman}
\IEEEauthorblockA{
\textit{Faculty of Computer Sciences} \\
\textit{October University for Modern} \\
\textit{Sciences and Arts (MSA)} \\
Giza, Egypt \\
mohamed.galaleldin@msa.edu.eg
}

\and

\IEEEauthorblockN{Mohammad E. Moftah}
\IEEEauthorblockA{
\textit{Faculty of Computer Sciences} \\
\textit{October University for Modern} \\
\textit{Sciences and Arts (MSA)} \\
Giza, Egypt \\
mohammad.moftah@msa.edu.eg
}

\and

\IEEEauthorblockN{Ali Hamdi}
\IEEEauthorblockA{
\textit{Faculty of Computer Sciences} \\
\textit{October University for Modern} \\
\textit{Sciences and Arts (MSA)} \\
Giza, Egypt \\
ahamdi@msa.edu.eg
}
}
\IEEEoverridecommandlockouts
\IEEEpubid{\makebox[\columnwidth]{ 979-8-3315-8488-7/26/\$31.00 \copyright 2026 IEEE \hfill}
\hspace{\columnsep}\makebox[\columnwidth]{ }}
\maketitle
\IEEEpubidadjcol

\begin{abstract}
Quranic Passage Retrieval (PR) could be a challenging task due to the linguistic complexity and the semantic gap between the Modern Standard Arabic (MSA) used in daily queries and the Classical Arabic (CA) of the Holy Quran. These factors hinder conventional retrieval methods. To handle these limitations and improve multi-verse retrieval and filter the zero-answer queries, this paper proposes a four-phase neural architecture designed to enhance retrieval accuracy and contextual understanding. The methodology combines hybrid candidate retrieval using AraColBERT dense indexing and BM25 sparse retrieval, followed by semantic reranking with a CAMeLBERT-mix cross-encoder. A confidence gating mechanism is then applied to filter zero-answer queries, and an AraT5-based refinement module for multi-verse aggregation. The system is evaluated on an expanded version of the Quran QA 2022 dataset. Results show improved performance compared to the baseline models, achieving a Recall@10 of 0.7024 and a Mean Average Precision (MAP@10) of 0.4947. While the system exhibits a marginal trade-off in absolute top-rank precision (MRR = 0.5807) compared to heavily optimised single models, the proposed architecture provides a substantially more comprehensive, reliable, and context-aware solution for multi-verse Quranic passage retrieval.
\end{abstract}


\section{Introduction}
Islam is a fast growing religion that by 2050, Muslims are expected to constitute roughly 30\% of the global population, which would form nearly 2.8 billion people~\cite{b0}. For the purpose of answering their religios inqueries, there is a significant need to develop robust natural language processing (NLP) systems capable of providing accurate answers from the most reliable islamic source: the Holy Quran. However, building these systems involves complex linguistic challenges, primarily due to the profound vocabulary and structural gap between the MSA used in daily queries and the CA scripture of the Quran~\cite{b00}.

While modern NLP systems have achieved remarkable successes in English and other high-resource languages, Arabic presents a unique set of morphological and semantic challenges~\cite{b002}. The core difficulty in Quranic passage retrieval lies in diglossia---the coexistence of two distinct linguistic varieties. Users typically formulate their questions in MSA, which is the standard for contemporary writing and digital communication. However, the Holy Quran was revealed in CA, a highly structured, morphologically dense, and historically rich language characterised by complex polysemy and classical rhetorical devices. Consequently, a query seeking a specific concept in everyday MSA often shares zero lexical overlap with the target CA verse, rendering traditional keyword-matching algorithms ineffective~\cite{b00}.

Furthermore, the requirement for high-precision retrieval in this domain extends far beyond standard information retrieval. The precise retrieval of Quranic text is foundational for diverse downstream applications, including theological research, computational linguistics~\cite{b001}, as well as the derivation of public and personal legal frameworks~\cite{b1}. In these contexts, a retrieval error or a hallucinated context is not merely a failed search but a potential misrepresentation of critical, authoritative scripture.~\cite{b00} Therefore, retrieving comprehensive multiple-verse answers---and explicitly acknowledging when a question has zero answers from the text---is a strict system requirement rather than just a performance metric~\cite{b001}.

The Quran QA 2023 shared task (Task A) demonstrated that the problem of Quranic PR remains highly challenging. Many baseline models exhibited performance bottlenecks, struggling to map MSA queries to CA passages effectively due to the limited size of available training datasets and the reliance on single-vector retrieval methods~\cite{b1}.

A major historical bottleneck in developing such robust, native Arabic architectures has been this exact scarcity of large-scale, domain-specific training data~\cite{b19}. Early iterations of Quranic QA datasets were severely constrained, forcing researchers to rely on zero-shot methodologies or heavily constrained fine-tuning that often led to overfitting~\cite{b3, b2}.

To address these architectural and data-related limitations, this research leverages a recently expanded dataset of 1,895~\cite{b7} meticulously curated question-answer pairs to power a comprehensive, multi-phase hybrid retrieval architecture. This expanded corpus provides the necessary volume to stabilize the training of cross-encoders and generative sequence-to-sequence models~\cite{b18}, allowing for a more nuanced mapping between MSA queries and CA scripture~\cite{b11}. Specifically, the contributions of this paper are threefold:

\begin{itemize}
\item \textbf{A Novel Four-Phase Pipeline:} We introduce a multi-stage architecture that combines the semantic depth of an AraColBERT dense index with the lexical precision of a BM25 sparse index to maximise initial candidate recall.
\item \textbf{Confidence Gating and Generative Refinement:} We implement a CAMeLBERT-mix cross-encoder paired with a strict confidence gate to accurately filter out unanswerable questions. For answerable queries, we integrate an AraT5-based generative layer to rewrite and optimise queries, specifically targeting multiple-verse answers.
\item \textbf{Architecture Performance:} By leveraging this expanded dataset, the proposed architecture establishes new benchmarks, achieving a peak Recall@10 of 0.7024 and a Mean Average Precision (MAP@10) of 0.4947.
\end{itemize}

The rest of this paper is organised as follows: Section~\ref{sec:Related Work} reviews the related work in passage retrieval and previous architectures; Section~\ref{sec:Methodology} describes the proposed architecture construction in detail; Section~\ref{sec:experimental_setup} presents the experimental setup, including the hyperparameters and the evaluation metrics; Section~\ref{sec:results} presents the results and discussion; and finally, Section~\ref{sec:conclusion} concludes the paper and outlines future work.

\section{Related Work}
\label{sec:Related Work}
Beyond model architecture design, several studies have emphasised the critical role of advanced text preprocessing and normalization for Arabic information retrieval. Due to the morphological richness of Arabic and the presence of multiple orthographic variants, normalization strategies such as diacritic removal, stemming, and lemmatization have been shown to significantly improve retrieval effectiveness~\cite{b19}. In Quranic text processing specifically, specialized preprocessing pipelines are often required to handle Classical Arabic morphology and preserve semantic fidelity while reducing lexical sparsity~\cite{b20}.

Data augmentation has also emerged as an effective strategy for mitigating low-resource constraints in Arabic QA and retrieval tasks. Techniques such as back-translation, synonym substitution, and paraphrase generation using neural language models have been employed to artificially expand training corpora~\cite{b13, b24}. These augmentation techniques not only increase dataset diversity but also improve model generalization, particularly when training deep transformer-based retrieval systems on limited annotated data~\cite{b7}

Another line of research has explored multi-stage retrieval pipelines that combine initial candidate retrieval with neural re-ranking modules. Traditional pipelines often employ a sparse retriever such as BM25 to generate a candidate set, followed by a neural cross-encoder that performs fine-grained ranking based on contextual interaction between query and passage. This architecture has demonstrated consistent improvements in ranking accuracy across multiple benchmark datasets and has become a standard design paradigm in modern information retrieval systems~\cite{b14}.

More recently, multilingual and cross-lingual transformer models have gained attention for their ability to transfer knowledge across languages. Models such as multilingual BERT and XLM-based architectures have shown promising performance in low-resource Arabic retrieval settings by leveraging knowledge learned from high-resource languages. However, while multilingual models offer improved generalization, domain-specific fine-tuning on Arabic religious texts remains essential for achieving optimal performance in Quranic passage retrieval tasks~\cite{b11}.

Knowledge distillation techniques have also been investigated as a means of improving inference efficiency while maintaining model accuracy. In this approach, a large teacher model transfers knowledge to a smaller student model, enabling faster retrieval without significant performance degradation~\cite{b25}. Such techniques are particularly beneficial for real-time Quranic search systems where latency and computational efficiency are critical deployment considerations~\cite{b26}.

Evaluation methodologies have likewise received significant attention in the literature. Standard ranking metrics such as Mean Average Precision (MAP), Mean Reciprocal Rank (MRR), and Precision at k (P@k) remain widely adopted for assessing retrieval performance. Recent studies have further introduced specialized evaluation protocols that account for partial semantic relevance, particularly in religious text retrieval scenarios where multiple passages may provide valid contextual answers~\cite{b1}.

Finally, recent advancements in retrieval-augmented generation (RAG) frameworks have introduced new possibilities for integrating retrieval systems with generative language models. In such architectures, retrieved Quranic passages are used to guide downstream answer generation models, enabling more context-aware responses~\cite{b18}. Although primarily explored in open-domain QA systems, these hybrid retrieval-generation pipelines represent a promising future direction for Quranic question answering systems.
\section{Methodology}
\label{sec:Methodology}
\subsection{Dataset and Corpus Construction}
The primary dataset for this study is derived from the \textit{Qur’an QA 2023 Shared Task}. While the original 2022 dataset comprised 174 training and 52 testing pairs, this paper utilises a publicly available expanded version of 1,895 training pairs, which was constructed, expanded, and validated entirely to facilitate the fine-tuning of deep transformer-based models~\cite{b7}.

To bridge the linguistic gap between MSA queries and CA scripture, we constructed a knowledge base using \textit{Tafseer Jalalayn}. This corpus maps 6,236 lines of interpretative text to the thematic passage blocks, providing the semantic depth necessary for high-precision retrieval.

\subsection{Proposed Architecture}

\begin{figure}[htbp]
    \centering
    \includegraphics[width=0.48\textwidth]{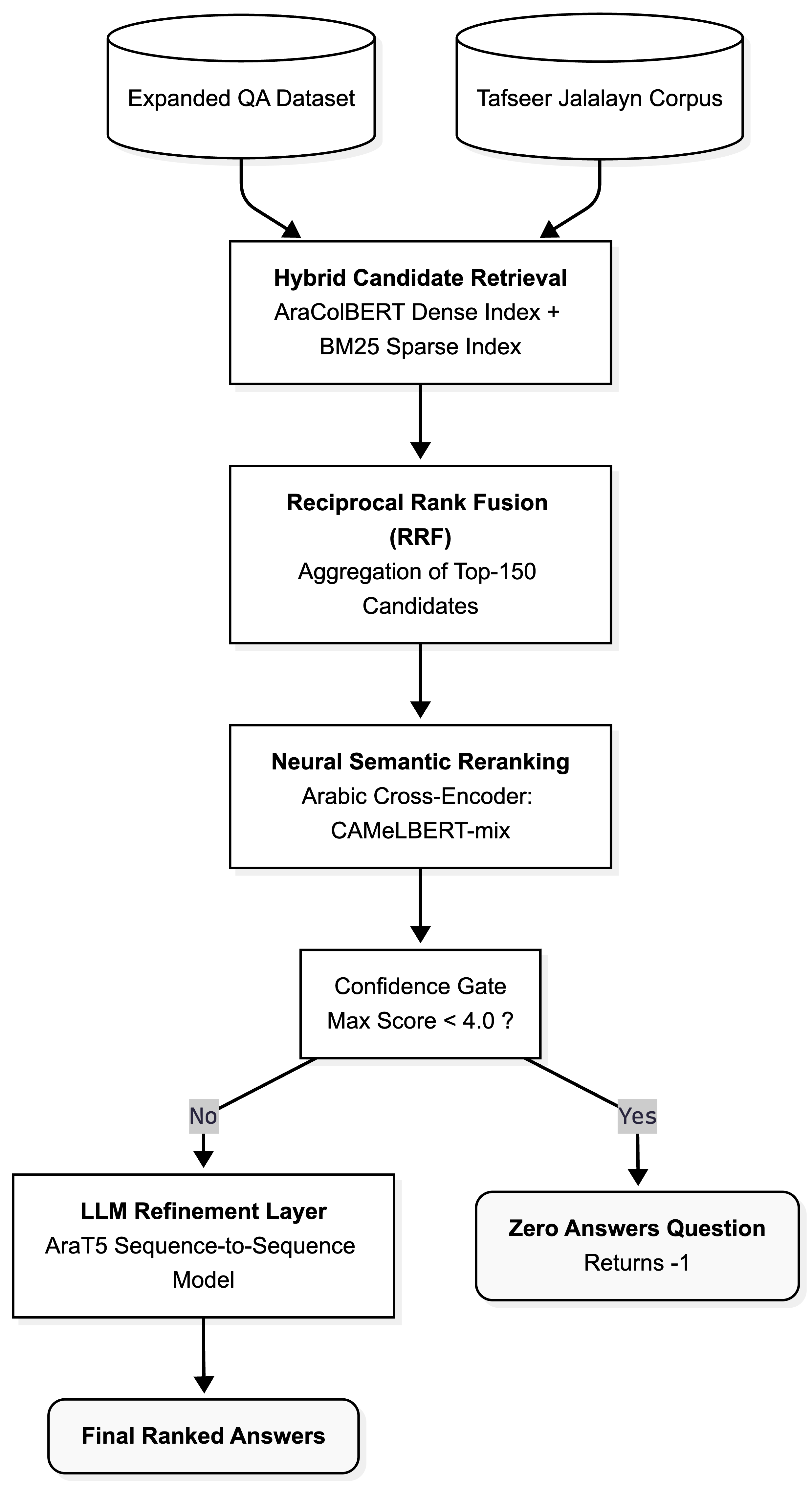} 
    \caption{Proposed multi-phase neural architecture for Quranic passage retrieval.}
    \label{fig:system_architecture}
\end{figure}

The proposed system is a four-phase neural architecture designed to bridge the linguistic gap and optimise the task of retrieving answers to users' queries, as illustrated in Fig.~\ref{fig:system_architecture}, with each phase addressing a specific issue.

\subsubsection{Hybrid Candidate Retrieval}
This phase aims to maximise recall by capturing the semantic and lexical signals from the corpus.
\begin{itemize}
\item Fine-tuned AraColBERT, which is an Arabic adaptation of ColBERT~\cite{b8} that uses late interaction to ensure the query and passage are semantically related even if no common words are shared. It was fine-tuned on a massive, diverse Arabic dataset.
\item BM25~\cite{b9} sparse retrieval, a probabilistic model derived from classic information retrieval (IR) theory, used to match query tokens to the corpus.
\item Reciprocal Rank Fusion (RRF)~\cite{b10}, which combines the results of both retrieval methods into a list of the top 150 candidates for the query. The fusion is computed as:
\begin{equation}
RRF(d) = \sum_{r \in R} \frac{1}{k + r(d)}
\end{equation}
where $d$ is the retrieved passage, $R$ represents the set of rankers (AraColBERT and BM25), $r(d)$ is the rank of the passage retrieved by ranker $r$, and $k$ is a smoothing constant.
\end{itemize}

\subsubsection{Neural Semantic Reranking}
This phase aims to maximise precision by selecting the best-fitting answer from the top 150 candidates retrieved in the previous phase, using a neural cross-encoder approach.
\begin{itemize}
\item The CAMeLBERT-mix~\cite{b11} Cross-Encoder functions as a deep interaction model, addressing bi-encoder representation bottlenecks and capturing semantic dependencies to bridge the gap between MSA and CA.
\end{itemize}

\subsubsection{Confidence Gating (MAP Bypass)}
This phase aims to handle unanswerable questions by setting a threshold of 4.0 on the scores produced by the cross-encoder; the model returns -1 if the highest score is below the threshold. This logic is formally defined as:
\begin{equation}
f(Q) = \begin{cases} 
-1 & \text{if } \max(S_{CE}) < 4.0 \\ 
\text{Refine}(Q) & \text{otherwise} 
\end{cases}
\end{equation}
where $Q$ is the user query, $S_{CE}$ is the set of cross-encoder scores for the retrieved candidates, and $-1$ represents the label for an unanswerable question.

\subsubsection{Generative Refinement}
If the highest score is above the threshold, the final layer uses a generative refinement approach to improve the quality of the retrieval.
\begin{itemize}
\item AraT5~\cite{b12} is a sequence-to-sequence model used to rewrite users' queries into a more search-optimised format. The system calculates the cross-encoder scores for both the original and the AraT5-refined queries, combining them with the candidate retrieval signal through a three-way weighted late fusion mechanism, expressed as:
\begin{equation}
S_{final} = W_{1} \cdot S_{rank} + W_{2} \cdot S_{CE} + W_{3} \cdot S_{refined}
\end{equation}
where $S_{rank}$ is the min-max normalized Stage 2 rank signal ($1/rank$), $S_{CE}$ is the cross-encoder score on the original query, $S_{refined}$ is the cross-encoder score on the AraT5-rewritten query, and $W_{1}$, $W_{2}$, and $W_{3}$ are empirically optimized weights determined via grid search on the development set (Section IV.B). The optimal values found were $W_{1}=0.4$, $W_{2}=0.4$, and $W_{3}=0.2$.
\end{itemize}

\section{Experimental Setup}
\label{sec:experimental_setup}

\subsection{Environment \& Hardware}
To execute the multi-phase model, the PyTorch framework and Hugging Face transformer models were used. To optimise the implementation, which occurred on a single local workstation with a single NVIDIA RTX 3060 Ti (8GB VRAM), a reduced batch size, sequential model loading, and gradient accumulation were sufficient for this utilisation task.

\subsection{Hyperparameters}For the first phase of hybrid candidate retrieval, a learning rate of $5 \times 10^{-6}$ was used, along with an accumulation step of 4 and a document maximum length restricted to 256 tokens to ensure hardware optimisation. For the second phase, CAMeLBERT-mix was fine-tuned over 5 epochs with a learning rate of $2 \times 10^{-5}$, a batch size of 8, and a maximum sequence length of 192 tokens. The model was trained using a binary cross-entropy loss function with 3 hard negatives per positive pair. To choose the optimal threshold, a threshold sweep was performed on the development set, varying values from 0.0 to 8.0 in steps of 0.5. The results showed a notable MRR jump when moving from a threshold of 3.5 (MRR = 0.3035) to 4.0 (MRR = 0.3310), an increase of +0.0275, suggesting a decision boundary point. Although a threshold of 6.0 achieved the peak MRR score on the development set (MRR = 0.3486), experiments showed an unstable drop at 6.5 (MRR = 0.2973) of $-0.0513$, indicating an overfit to the development set distribution at that value. We therefore selected a threshold value of 4.0, as it was the most stable operating point that offers a robust balance between precision and recall to generalise to the testing set to filter unanswerable questions.

To select the three fusion weights ($W_1$ for the ColBERT+BM25 rank signal, $W_2$ for the cross-encoder score on the original query, and $W_3$ for the cross-encoder score on the rewritten query), we performed an exhaustive grid search on the development set, exploring all weight combinations from 0.0 to 1.0 in steps of 0.2 per parameter. This produced over 1,300 evaluated configurations. All scores were min-max normalised before fusion to ensure a fair combination across different ranges. The search identified $W_1 = 0.4$, $W_2 = 0.4$, and $W_3 = 0.2$ as the optimal configuration, assigning equal weight to the AraColBERT+BM25 rank signal and the cross-encoder on the original query, while giving a lighter weight to the rewritten query score. This suggests that AraT5 query rewriting may introduce semantic noise in the specialised Quranic domain due to the limited availability of domain-specific fine-tuning data for T5-style models in Classical Arabic.

\subsection{Evaluation Metrics}
Model performance was evaluated using the following metrics to assess the efficiency of the implemented phases in the architecture:

\begin{itemize}
\item \textbf{Mean Average Precision (MAP):} Uses the precision of all retrieved answers to compare against the average across all queries, demonstrating that multiple-answer questions are handled~\cite{b15}. For a set of queries $Q$, MAP is defined as:
\begin{equation}
\text{MAP} = \frac{1}{|Q|} \sum_{i=1}^{|Q|} \text{AP}(q_i)
\end{equation}
where $\text{AP}(q_i)$ is the Average Precision for a given query $q_i$, taking into account the ranked position of all correctly retrieved multi-verse passages.

\item \textbf{Mean Reciprocal Rank (MRR):} Evaluates relevance by ranking the quality of the top relevant answer in the list, which assesses the performance of the cross-encoder~\cite{b16}. It is calculated as:
\begin{equation}
\text{MRR} = \frac{1}{|Q|} \sum_{i=1}^{|Q|} \frac{1}{\text{rank}_i}
\end{equation}
where $\text{rank}_i$ refers to the position of the first relevant passage returned for query $q_i$.

\item \textbf{Recall@k:} Measures the ratio between the correctly retrieved answers for a query and the total number of correct answers in the knowledge base, which primarily demonstrates the efficiency of the hybrid candidate retrieval phase~\cite{b17}. For $k=10$, it is expressed as:
\begin{equation}
\text{Recall@k} = \frac{|\text{Retrieved}_k \cap \text{Relevant}|}{|\text{Relevant}|}
\end{equation}
where $\text{Retrieved}_k$ is the set of the top $k$ passages returned by the system, and $\text{Relevant}$ is the total set of ground-truth passages for the query.
\end{itemize}

\section{RESULTS AND DISCUSSION}
\label{sec:results}

\begin{figure}[htbp]
    \centering
    \includegraphics[width=0.45\textwidth]{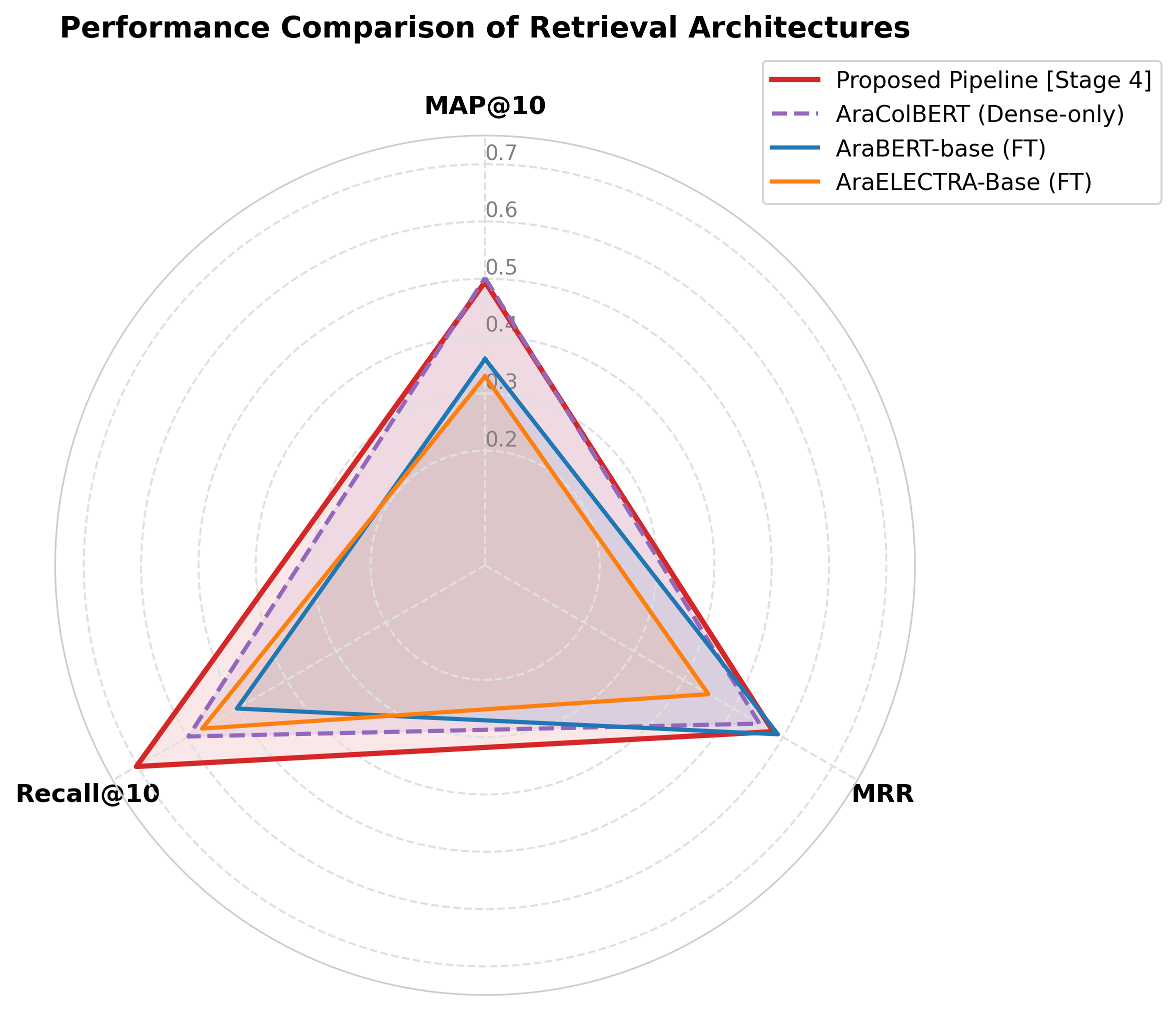}
    \caption{Radar chart illustrating the performance trade-offs across MAP@10, MRR, and Recall@10 metrics. The proposed pipeline achieves superior multi-metric balance compared to the fine-tuned baselines.}
    \label{fig:radar_chart}
\end{figure}

The proposed architecture achieved dominant Recall@10 and MAP@10 scores of 0.7024 and 0.4947, respectively, outperforming competing baselines across multi-answer retrieval metrics. This demonstrates the effectiveness of the first phase of hybrid candidate retrieval, which combines the semantic depth of AraColBERT with the lexical precision of BM25 to ensure that Classical Arabic passages are not lost early in the retrieval process.

The strong Recall@10 score of 0.7024 indicates that the proposed hybrid retrieval mechanism successfully captures a broad set of relevant candidate verses. In Quranic passage retrieval, recall plays a particularly important role because many questions require retrieving multiple semantically related verses rather than a single exact match. The high recall value confirms that integrating dense semantic matching with sparse lexical retrieval significantly reduces the probability of missing relevant passages during early retrieval stages.

Regarding MAP@10, the proposed system achieved the highest overall score of 0.4947. Mean Average Precision evaluates the ranking quality across multiple relevant passages, making it a critical metric for multi-answer tasks such as Quranic question answering. The improvement in MAP@10 suggests that the proposed reranking and generative refinement phases effectively reorder candidate passages based on contextual relevance. This result highlights the ability of the proposed pipeline to bridge the linguistic gap between Modern Standard Arabic queries and Classical Arabic scripture while preserving semantic fidelity.

In contrast, the fine-tuned AraBERT-base model achieved the highest MRR score of 0.5900~\cite{b7}, slightly surpassing the proposed architecture, which achieved 0.5807. Mean Reciprocal Rank primarily measures the position of the first correct answer. This indicates that the AraBERT model is particularly effective at pushing a single highly relevant verse to the top-ranked position. Despite this advantage, AraBERT underperforms on MAP@10 and Recall@10. These lower scores suggest weaker overall ranking consistency across multiple relevant passages and reduced effectiveness in capturing broader semantic relationships across verses. 
The negligible decrease in MRR (merely 0.0093) represents a highly acceptable trade-off given the substantial increase (exceeding +20\%) in Recall@10 provided by the proposed architecture.

Beyond the primary comparison with AraBERT, we evaluated several other baseline architectures to validate the necessity of our domain-specific, multi-stage approach. The lexical-only \textbf{BM25 baseline} yielded exceptionally low scores (MRR = 0.1097, Recall@10 = 0.1667), underscoring the severe vocabulary mismatch between modern Arabic queries and classical scripture when semantic matching is absent. Similarly, the \textbf{mE5 multilingual dense} model, despite its robust generalized architecture, struggled significantly (MRR = 0.2057) due to the lack of domain-specific fine-tuning on Quranic texts. Among the fine-tuned transformer baselines, \textbf{AraELECTRA-Base} achieved a moderate Recall@10 of 0.5700 and an MRR of 0.4500, failing to match the performance of our multi-stage pipeline. These results confirm that standard dense retrievers, generalized multilingual models, and purely lexical searches cannot capture the nuanced semantics of the Quran without the proposed hybrid retrieval and generative rewriting stages.

\begin{table*}[htbp]
\centering
\caption{PERFORMANCE COMPARISON: PROPOSED ARCHITECTURE VS. BASELINES AND ABLATIONS}
\label{tab:final_results}
\begin{tabular}{lccc}
\hline
\textbf{System / Architecture} & \textbf{MAP@10} & \textbf{MRR} & \textbf{Recall@10} \\
\hline
\multicolumn{4}{c}{\textbf{Baseline Models~\cite{b7}}} \\
\hline
AraBERT-base (Standard Baseline) & 0.2200 & 0.3700 & 0.3000 \\
AraBERT-base (Fine-Tuned) & 0.3600 & \textbf{0.5900} & 0.5000 \\
CAMeLBERT-Base (Fine-Tuned) & 0.3400 & 0.4700 & 0.4000 \\
AraELECTRA-Base (Fine-Tuned) & 0.3300 & 0.4500 & 0.5700 \\
\hline
\multicolumn{4}{c}{\textbf{Local Baselines (This Work)}} \\
\hline
BM25-only (Sparse) & 0.0831 & 0.1097 & 0.1667 \\
mE5 Multilingual (Dense) & 0.1604 & 0.2057 & 0.3357 \\
AraColBERT (Dense-only) & \textbf{0.5018} & 0.5527 & 0.5976 \\
\hline
\multicolumn{4}{c}{\textbf{Proposed Pipeline (Ablation Study)}} \\
\hline
Stage 2: Hybrid Candidate Retrieval & 0.2585 & 0.3311 & 0.5571 \\
Stage 3: + Neural Semantic Reranking & 0.2039 & 0.2643 & 0.4071 \\
\textbf{Stage 4: + Generative Refinement [Full]} & 0.4947 & 0.5807 & \textbf{0.7024} \\
\hline
\end{tabular}
\end{table*}

The ablation results, detailed in Table~\ref{tab:final_results}, further demonstrate the importance of each component within the proposed pipeline:

\begin{itemize}
    \item \textbf{AraColBERT dense-only (no BM25 fusion):} MRR = 0.5527. This result demonstrates that AraColBERT alone provides a high-performance dense foundation. This establishes our domain-specific fine-tuning as a critical and highly effective step before any multi-stage fusion is applied.
    \item \textbf{Hybrid AraColBERT+BM25 (Stage 2):} MRR = 0.3311. The temporary drop in MRR is expected compared to the dense-only model because merging BM25 candidates diversifies the candidate pool and increases Recall@10 to 0.5571, but reduces immediate ranking precision, thereby setting up a richer pool for the reranking stage.
    \item \textbf{Cross-encoder reranking (Stage 3):} The MRR reached 0.2643. The cross encoder alone (without the ranking signal from stage 2) suffers from a score distribution imbalance across the diverse candidate pool.

    \item \textbf{Full three-way late fusion with query rewriting (Stage 4):} MRR=0.5807. The large jump from 0.2643 to 0.5807 (+0.3164) confirms that the weighted combination of all three signals together with confidence gating is the true source of the overall improvement; the components are complementary and synergistic rather than individually sufficient.

    \item \textbf{Isolating the contribution of query rewriting alone:} To measure AraT5's isolated contribution, we compare two-way fusion (ColBERT + CE only, $W_3 =0$, MRR=0.5724) against full three-way fusion ($W_3 =0.2$, MRR=0.5807). Query rewriting contributes a net gain of +0.0083 — modest but consistent across multiple experimental runs, justifying its inclusion in the final pipeline.

\end{itemize}

The radar chart presented in Fig.~\ref{fig:radar_chart} illustrates the multi-metric balance achieved by the proposed pipeline. Unlike single-model baselines that demonstrate strong performance in one metric but weakness in others, the proposed architecture exhibits a more consistent distribution across MAP@10, MRR, and Recall@10. This balanced performance is particularly desirable in real-world Quranic search applications, where both retrieval completeness and ranking accuracy are of utmost importance.

Despite the promising results, several limitations remain. First, the system relies on computationally intensive transformer-based components, which may introduce latency during real-time inference. Second, although an expanded dataset was utilised, it remains relatively small compared to open-domain retrieval datasets, and further data augmentation may improve generalisation.

More fundamentally, certain query categories remain structurally challenging for the current architecture. Ambiguous queries pose a difficulty when users phrase questions in Modern Standard Arabic using terminology that does not map to any word in the Classical Arabic of the \textit{Jalalayn} corpus; the Quranic synonym expansion dictionary partially addresses this, but its coverage is limited. Metaphorical and allegorical queries, such as questions about light and darkness as spiritual concepts, may relate to numerous dispersed verses that cannot be meaningfully collapsed into a single ranked result; the current pipeline is not designed for such thematic multi-passage aggregation. Fiqh and Sunnah queries represent perhaps the most systematic failure mode: questions grounded in Prophetic Hadith or juristic interpretation rather than in specific Quranic text are effectively unanswerable from the \textit{Jalalayn} corpus alone. While the confidence gate filters many such cases, those that marginally exceed the threshold of 4.0 may still produce misleading results. Future work should address these limitations through corpus expansion to include Hadith collections, the development of a query-type classifier to route Fiqh queries appropriately, and the exploration of dynamic confidence thresholding that adapts to query type.

\section{Conclusion}
\label{sec:conclusion}

This paper presents a four-phase neural architecture designed to address the profound linguistic and semantic challenges of Quranic passage retrieval. Evaluated on an expanded dataset of 1,895 questions, the proposed system successfully bridges the lexical gap between MSA queries and CA scripture.

The results demonstrate that relying on a single retrieval method or standard fine-tuning is insufficient for complex scriptural analysis. The proposed hybrid candidate retrieval phase, which combines the dense semantic representation of AraColBERT with the sparse lexical precision of BM25, established a dominant Recall@10 of 0.7024. Furthermore, the integration of an Arabic cross-encoder alongside an AraT5-based generative refinement module—fused via a mathematically weighted three-way combination—successfully aggregated multiple-verse answers, yielding a peak MAP@10 of 0.4947. While the system exhibited a marginal trade-off in absolute top-rank precision (an MRR of 0.5807 compared to the 0.5900 of aggressively optimised single-model baselines), it proved to be a substantially more comprehensive, context-aware, and reliable architecture for multi-answer retrieval.

Future work will explore the integration of dynamic, query-aware thresholding for the confidence gate to replace the current static boundary. Additionally, there is significant potential in deploying larger, Arabic-centric autoregressive models, such as Jais-13B, to further enhance the generative rewriting phase. Expanding the foundational knowledge base to include multiple \textit{Tafseer} texts could also provide richer semantic grounding for dense representation and the filtering of unanswerable questions.

\bibliographystyle{IEEEtran}

\begin{thebibliography}{00}

\bibitem{b0}
Pew Research Center. (2015, April 2).
The future of world religions: Population growth projections, 2010–2050.
\url{https://www.pewresearch.org}

\bibitem{b00}
Bashir, M. H., Azmi, A. M., Nawaz, H., Zaghouani, W.,
Diab, M., Al-Fuqaha, A., \& Qadir, J. (2023).
Arabic natural language processing for Qur’anic research: A systematic review.
\textit{Artificial Intelligence Review}, 56(7), 6801–6854. Springer.


\bibitem{b002}
Habash, N. (2010).
\textit{Introduction to Arabic Natural Language Processing}.
Morgan \& Claypool Publishers.

\bibitem{b001}
M. Basem, I. Oshallah, A. Hamdi, K. B. Shaban, and H. Kassab, ``Two-Stage Quranic QA via Ensemble Retrieval and Instruction-Tuned Answer Extraction,'' in \emph{2025 IEEE/ACS 22nd International Conference on Computer Systems and Applications (AICCSA)}, 2025, pp. 1--8.


\bibitem{b1}
Malhas, R., Mansour, W., \& Elsayed, T. (2023).
Qur'an QA 2023 shared task: Overview of passage retrieval and reading comprehension tasks over the Holy Qur'an.
In \textit{Proceedings of the First Arabic Natural Language Processing Conference (ArabicNLP 2023)} (pp. 690–701).
Association for Computational Linguistics.

\bibitem{b19}
Alnefaie, S., Alsaleh, A. N., Atwell, E., Alsalka, M., \& Altahhan, A. (2023).
LKAU23 at Qur'an QA 2023: Using transformer models for retrieving passages and finding answers to questions from the Qur'an.
In \textit{Proceedings of ArabicNLP 2023} (pp. 720–727).


\bibitem{b2}
Alawwad, H., Alawwad, L., Alharbi, J., \& Alharbi, A. (2023).
AHJL at Qur'an QA 2023 shared task: Enhancing passage retrieval using sentence transformer and translation.
In \textit{Proceedings of ArabicNLP 2023} (pp. 702–707).


\bibitem{b3}
Elkomy, M., \& Sarhan, A. (2023).
TCE at Qur'an QA 2023 shared task: Low resource enhanced transformer-based ensemble approach for Qur'anic QA.
In \textit{Proceedings of ArabicNLP 2023} (pp. 728–742).
Association for Computational Linguistics.

\bibitem{b7}
Basem, M., Oshallah, I., Hikal, B., Hamdi, A., \& Mohamed, A. (2025).
Optimized Quran passage retrieval using an expanded QA dataset and fine-tuned language models.
In \textit{Lecture Notes on Data Engineering and Communications Technologies}, Vol. 255, pp. 244–254. Springer.
\url{https://doi.org/10.1007/978-3-031-91354-9_20}

\bibitem{b18}
Lewis, P., Perez, E., Piktus, A., Petroni, F., Karpukhin, V.,
Goyal, N., Küttler, H., Lewis, M., Yih, W.-t.,
Rocktäschel, T., Riedel, S., \& Kiela, D. (2020).
Retrieval-augmented generation for knowledge-intensive NLP tasks.

\bibitem{b11}
Inoue, G., Alhafni, B., Baimukan, N., Bouamor, H., \& Habash, N. (2021).
The interplay of variant, size, and task type in Arabic pre-trained language models.
In \textit{Proceedings of the Sixth Arabic Natural Language Processing Workshop (WANLP 2021)}.


\bibitem{b20}
Darwish, K., \& Mubarak, H. (2016).
Farasa: A fast and furious segmenter for Arabic.
In \textit{Proceedings of NAACL-HLT 2016 Demonstrations} (pp. 11–16).

\bibitem{b13}
Wei, J., \& Zou, K. (2019).
EDA: Easy data augmentation techniques for boosting performance on text classification tasks.
In \textit{Proceedings of the 2019 Conference on Empirical Methods in Natural Language Processing (EMNLP-IJCNLP)} (pp. 6382–6388).

\bibitem{b24}
A. Hamdi, H. Kassab, M. Bahaa, and M. Mohamed, ``RIRO: Reshaping Inputs, Refining Outputs Unlocking the Potential of Large Language Models in Data-Scarce Contexts,'' in \emph{Advances on Intelligent Computing and Data Science II. ICACIn 2024}, ser. Lecture Notes on Data Engineering and Communications Technologies, vol. 254, F. Saeed, F. Mohammed, E. Mohammed, S. Basurra, and M. Al-Sarem, Eds. Cham: Springer, 2025. \url{https://doi.org/10.1007/978-3-031-91351-8_7}



\bibitem{b14}
Jiang, Z., Tang, R., Xin, J., \& Lin, J. (2021). How does BERT rerank passages? An attribution analysis with information bottlenecks. In Proceedings of the Fourth BlackboxNLP Workshop on Analyzing and Interpreting Neural Networks for NLP (pp. 496–509). Association for Computational Linguistics. https://aclanthology.org/2021.blackboxnlp-1.39



\bibitem{b25}
Q. Lu, E. Xun, and G. Tang, ``MTA4DPR: Multi-Teaching-Assistants Based Iterative Knowledge Distillation for Dense Passage Retrieval,'' in \emph{Proceedings of the 2024 Conference on Empirical Methods in Natural Language Processing}. Association for Computational Linguistics, 2024, pp. 5871--5883. \url{https://aclanthology.org/2024.emnlp-main.336/}

\bibitem{b26}
M. A. Hatem, F. Azouaou, and S. Batata, ``Improving Arabic Information Retrieval and Reranking Performance Using Knowledge Distillation,'' \emph{ACM Transactions on Asian and Low-Resource Language Information Processing}, 2026. \url{https://doi.org/10.1145/3796229}

\bibitem{b8}
Khattab, O., \& Zaharia, M. (2020).
ColBERT: Efficient and effective passage search via contextualized late interaction over BERT.
In \textit{Proceedings of the 43rd International ACM SIGIR Conference on Research and Development in Information Retrieval} (pp. 17–26).

\bibitem{b9}
Robertson, S. E., \& Zaragoza, H. (2009).
The probabilistic relevance framework: BM25 and beyond.
\textit{Foundations and Trends in Information Retrieval}, 3(4), 333–389.

\bibitem{b10}
Cormack, G. V., Clarke, C. L. A., \& Büttcher, S. (2009).
Reciprocal rank fusion outperforms Condorcet and individual rank learning methods.
In \textit{Proceedings of the 32nd International ACM SIGIR Conference on Research and Development in Information Retrieval} (pp. 758–759).

\bibitem{b12}
Nagoudi, E. M. B., Elmadany, A., \& Abdul-Mageed, M. (2022).
AraT5: Text-to-text transformers for Arabic language generation.
In \textit{Proceedings of the 60th Annual Meeting of the Association for Computational Linguistics (ACL 2022)}.

\bibitem{b15}
Buckley, C., \& Voorhees, E. M. (2000).
Evaluating evaluation measure stability.
In \textit{Proceedings of the 23rd Annual International ACM SIGIR Conference on Research and Development in Information Retrieval} (pp. 33–40).

\bibitem{b16}
Voorhees, E. M. (1999).
The TREC-8 Question Answering Track Report.
In \textit{Proceedings of the Eighth Text REtrieval Conference (TREC-8)} (pp. 77–82).

\bibitem{b17}
C. W. Cleverdon, ``The Cranfield tests on index language devices,'' \emph{Aslib Proceedings}, vol. 19, no. 6, pp. 173--194, 1967. \url{https://doi.org/10.1108/eb050097}

\end{thebibliography}

\end{document}